\documentstyle[twoside,fleqn,espcrc2]{article}
\renewcommand{\Re}{\mbox{Re}\,}
\renewcommand{\Im}{\mbox{Im}\,}

\newcommand{\AmS}{{\protect\the\textfont2
  A\kern-.1667em\lower.5ex\hbox{M}\kern-.125emS}}
\title{
{\small NYU-TH-96/09/01 hep-th/9609073}
\\ \vskip .1in
Spontaneous Breaking of Extended Supersymmetry in Global and Local
Theories
\thanks{Contribution to the proceedings of the {\em
Spring School and Workshop on String Theory, Gauge Theory and Quantum
Gravity}, ICTP, Trieste, Italy, March 18-29, 1996.}
}
\author{M. Porrati\address{Department of Physics, NYU \\
        4 Washington Pl. New York, NY 10003, USA}
        \thanks{Supported in part by NSF under grant PHY-9318781}}

\begin{document}

\begin{abstract}
We review the ``no-go'' theorems that severely constrain the breaking of
$N=2$ supersymmetry to $N=1$ (both in rigid supersymmetry and supergravity),
and we exhibit some models that evade them.
\end{abstract}

\maketitle

\section{A Theorem and its Loophole}
A well known theorem~\cite{W} ensures that extended rigid
supersymmetry cannot be broken partially: either all supersymmetries are
broken or all are exact. The heuristic proof of this theorem follows from
the $N$-extended SUSY algebra:
\begin{equation}
\{ \bar{Q}_{\dot{\alpha}}^i,
Q_\alpha^j\}=2\sigma^\mu_{\dot{\alpha}\alpha}\delta^{ij}P_\mu, \;\;
i=1,..,N.
\label{1}
\end{equation}
The Hamiltonian of a supersymmetric theory is manifestly positive, because,
thanks to eq.~(\ref{1}), it can be expressed as
\begin{equation}
H={1\over 2}\sum_\alpha Q^{\dagger\, i}_\alpha Q^i_\alpha, \;\; \mbox{no sum
over $i$}.
\label{2}
\end{equation}
If a supersymmetry, say $i=1$, is broken,
then its generator does not annihilate the
vacuum: $Q^1|0\rangle\neq 0$. Equation~(\ref{2}) then implies that $\langle
0|H|0\rangle\neq 0$, i.e. that $Q^i|0\rangle\neq 0$ for all $i$.
This argument can be made rigorous either by working with a theory in a
finite periodic box (thereby giving up Lorentz invariance), or by using the
SUSY current algebra
\begin{eqnarray}
\lim_{V\rightarrow \infty}\{\bar{Q}^i_{V\,\dot{\alpha}},J_{\mu\,
\alpha}^j(x)\}
&=& 2\sigma^\nu_{\dot{\alpha}\alpha}\delta^{ij}T_{\mu\nu}(x),\nonumber \\
\bar{Q}^i_{V\, \dot{\alpha}} &\equiv & \int_V d^3\vec{y}\bar{J}^i_{0\,
\dot{\alpha}}(\vec{y},t).
\label{3}
\end{eqnarray}
Here $T_{\mu\nu}$ and $J_{\mu\, \alpha}^i$ are the stress-energy tensor and
the supercurrent, respectively. The infinite volume limit in this equation
exists even when SUSY is broken\footnotemark,
\footnotetext{That is when, rigorously speaking,
the supercharge does not exist as an operator in the Hilbert space.}
because
the (anti)-commutator of local (gauge-invariant) operators is always local.

The strength of SUSY breaking
is parametrized by the matrix element of the SUSY current in between the
vacuum and the (would-be) goldstino states
\begin{equation}
\langle p\xi j| J^i_{\mu\, \alpha}(x)|0\rangle = \bar{v}^{\dot{\alpha}\,
j}\sigma_{\mu\, \dot{\alpha}\alpha}F^i_j e^{ipx}.
\label{4}
\end{equation}
By computing the vacuum expectation value of eq.~(\ref{3}), and using
eq.~(\ref{4}), one finds that $F^{*\, i}_l F^l_j \propto \delta^i_j \langle
0| T_\mu^\mu |0\rangle$. The SUSY algebra has an $SU(N)$ symmetry that acts
by unitary transformations on the index $i$, and that can be used to bring
the goldstino-coupling matrix into the form $F^i_j=F_g\delta^i_j$.
This shows that either $F_g=0$, i.e. all supersymmetries are exact, or
$F_g\neq 0$, that is they are all broken with equal strength.

This ``no-go'' theorem can be evaded, as often goes with no-goes, by
changing its assumptions. They can be changed in two ways
\begin{enumerate}
\item If the Hilbert space of the theory is not positive definite, then
$Q^i|0\rangle\neq 0$ can be consistent with
$\langle 0|Q^{\dagger \,i}Q^i|0\rangle= 0$.
This is what happens in covariant formulations of supergravity.
\item The SUSY current algebra is modified. This happens in supergravity
when the local supersymmetry is gauge-fixed non-covariantly. More
surprisingly, the SUSY current algebra may be modified even in rigid
supersymmetry~\cite{HLP,FGP}.
\end{enumerate}
In the latter case the modification reads
\begin{eqnarray}
\lim_{V\rightarrow \infty}\{\bar{Q}^i_{V\,\dot{\alpha}},J_{\mu\,
\alpha}^j(x)\}
&=& 2\sigma^\nu_{\dot{\alpha}\alpha}\delta^{ij}T_{\mu\nu}(x)+ \nonumber \\ &&
+\sigma_{\mu\,\dot{\alpha}\alpha}C^{ij}. \label{5}
\end{eqnarray}
The additional term $C^{ij}$ is a constant, thus, the supersymmetry algebra
{\em on local operators}
is not modified by its presence. By studying the
SUSY algebra on local operators only, one would never be able to
detect the presence of  $C^{ij}$, yet this term is enough to invalidate the
``no-go'' theorem!

The question remains as to whether such a constant nonzero term can arise in
a local field theory with rigid $N=2$ supersymmetry.

The answer was given in ref.~\cite{APT}. \cite{APT} explicitly construct an
$N=2$ theory (hereafter called the APT model),
with nonzero $C^{ij}$. The field content of the theory is a
single $N=2$ vector multiplet, whose helicity content
is $[(1)\pm 1,(2)\pm 1/2 ,(2)0]$.

The $N=2$ Lagrangian of the vector multiplet is completely determined by an
analytic function (the prepotential) ${\cal F}(z)$
($z$ is the complex scalar in the vector multiplet), and by two constant
vectors of the $SU(2)$ isomorphism group of the $N=2$ superalgebra,
$\vec{E}$, $\vec{M}$~\cite{APT}. $\vec{E}$ is the standard (electric)
$N=2$ Fayet-Iliopoulos term, while the magnetic F-I term $\vec{M}$
was introduced first
in~\cite{APT}. With an $SU(2)$ rotation they can be cast in the form
\begin{equation}
\Re \vec{E}=\Lambda^2 (0,e,\xi),\;\; \vec{M}=\Lambda^2(0,m,0).
\label{5a}
\end{equation}
The kinetic term of all fields in the multiplet is given by
$\tau_2(z)=\Im {\cal F}''(z)>0$, while the theta angle is $\tau_1(z)=\Re {\cal
F}''(z)$.

To find whether the $N=2$ supersymmetry is broken to $N=1$ one must look for
Lorentz-invariant backgrounds on which the SUSY transformation laws of some
field have a nonzero VEV. Lorentz invariance constrains the form of
a non-zero VEV to
\begin{equation}
\delta \lambda_i = {i\over
\sqrt{2}}Y^A\epsilon_{ij}\sigma^{A\,j}_k\eta^k.
\label{6}
\end{equation}
Here, $\lambda_i$ is the gaugino field (doublet of $SU(2)$), $\eta^i$ is
the $N=2$ SUSY transformation parameter, and $Y^A$, $A=1,2,3$, is
a yet unspecified (complex) vector of $SU(2)$. In ref.~\cite{APT} it was
shown that
\begin{equation}
Y^A=-{2\over \tau_2(z)} \left[ \Re E^A +\tau_1(z) M^A\right] +2iM^A.
\label{7}
\end{equation}

Partial breaking of $N=2$ to $N=1$ occurs if the matrix
$Y^A(a)\epsilon_{ij}\sigma^{A\, j}_k$ has a non-degenerate zero eigenvalue.
This happens at
\begin{equation}
\tau_1(z)=-e/m,\;\; \tau_2(z)=m/\xi.
\label{8}
\end{equation}
Generically,
these equations can be solved for $z$ if $m/\xi>0$, and ${\cal F}''(a)$
is not a constant, i.e. when the model is non-renormalizable and the
magnetic F-I term $\vec{M}$ is nonzero.
A VEV $z^o$ that solves eq.~(\ref{8}) is also a stable minimum of the
scalar potential. This is a general result, valid whenever some
supersymmetry is unbroken. It follows from the current algebra
eq.~(\ref{5}), which implies the following identity for the scalar potential
\begin{equation}
(\delta^i\lambda^I)^* Z_I^J(\phi)\delta^j\lambda_J= \delta^{ij}V(\phi) +
C^{ij}.
\label{9}
\end{equation}
Here $\delta^i\lambda_I$ is the $i$-th SUSY transformation law of the
spin-1/2 fields $\lambda_I$ (on a Lorentz-invariant background). The
fermion kinetic term is $Z_I^J(\phi)$, $\phi$
denotes all scalar fields; $V(\phi)$ is the potential. If a supersymmetry,
say $i=1$, is unbroken at $\phi=\phi^o$,
then $\delta^1\lambda_I=0$ $\forall I$, and
the component $11$ of the matrix equation above implies
\begin{equation}
V(\phi^o)=-C^{11},\;\; \left.{\partial V\over
\partial \phi}\right|_{\phi^o}=0,\;\; \left.{\partial^2 V\over
\partial \phi^2}\right|_{\phi^o} \geq 0.
\label{10}
\end{equation}

Explicitly computing the current algebra of the APT model one finds that the
field-independent term $C^{ij}$ is~\cite{FGP}
\begin{equation}
C^{ij}= 4\epsilon_{ABC}\sigma^{A\, ij} \Re E^B M^C .
\label{11}
\end{equation}
As expected, partial SUSY breaking in the APT model is possible only when
$C^{ij}\neq 0$.

Since the vacua given in eq.~(\ref{8}) are  $N=1$ supersymmetric, the $N=2$
vector multiplet decomposes into $N=1$ multiplets, specifically, into a
massless vector multiplet and a massive chiral multiplet.
\section{Supergravity and Partial Breaking}
In supergravity theories, there is no simple general argument forbidding
partial breaking, not even at zero cosmological constant. Some constraints
nevertheless exist.

To be concrete, let us examine the case of $N=2$ supergravity.
First of all, the minimal field content of an $N=2$ theory that may allow
for partial breaking to $N=1$ is, besides the $N=2$ gravitational multiplet,
a vector multiplet {\em and} a hypermultiplet~\cite{FV}.
The helicity content of the gravitational multiplet is $[(1) \pm 2, (2) \pm
3/2, (1) \pm 1]$; the content of the hypermultiplet is $[(2) \pm 1/2, (4)
0]$. The vector multiplet is necessary because partial breaking to $N=1$
requires that one of the spin-3/2 fields becomes massive. A massive spin-3/2
multiplet of $N=1$ contains two massive vectors, while the pure
gravitational multiplet of $N=2$ contains only one such field. Since the
scalars in the vector multiplet are neutral under the $U(1)$'s gauged
by the spin-1 fields, a
(charged) hypermultiplet is needed to give mass to the vectors by the Higgs
effect.

A more technical constraint follows from the details of the construction of
$N=2$ supergravity.

In an $N=2$ supergravity, the scalars in the vector multiplet, $z^a$,
$a=1,.., n_V$, parametrize a ``special K\"ahler'' manifold, i.e. a K\"ahler
manifold endowed with a symplectic bundle of dimension $2n_V+2$ which
possesses a holomorphic section $Z^\alpha(z)$, $\alpha=1,..,2n_V+2$.
The K\"ahler
potential is completely determined by the section as~\cite{CDF}
\begin{equation}
K=-\log i \bar{Z}^\alpha(z)\omega_{\alpha\beta}Z^\beta(z),
\label{12}
\end{equation}
where $\omega_{\alpha\beta}$ is the (constant) symplectic norm of the bundle.
{\em Symplectic transformations on the bundle also act as electric-magnetic
dualities on the $n_V+1$ vector fields} ( the matter vectors plus the
graviphoton).

The hypermultiplets, on the other hand, live in a quaternionic
manifold~\cite{BW}. They can be charged under the $n_V+1$ vectors. When
these vectors are Abelian,  the most
general charge allowed by the DSZ quantization conditions~\cite{DSZ} is a
$2n_V+2$-dimensional vector $q^\alpha_I$  obeying
$q^\alpha_I\omega_{\alpha\beta}q^\beta_J\in Z$,
with $I$ labelling the different hypermultiplets.
To be able to describe the interactions of the hypermultiplets with a
local Lagrangian, the charges must obey a stronger condition:
\begin{equation}
q^\alpha_I\omega_{\alpha\beta}q^\beta_J=0,\;\; \forall I,J.
\label{13}
\end{equation}
Obviously, there are at most $n_V+1$ linearly independent $q^\alpha_I$ obeying
such condition. Whenever eq.~(\ref{13}) holds, one can choose $n_V +1$
linearly independent vectors $c^{A\alpha}$,
$A=0,..,n_V$, obeying $q^\alpha_I\omega_{\alpha\beta}c^{A\beta}=0$. These
vectors select a particular symplectic basis with coordinates
\begin{equation}
X^A= c^{A\alpha}Z^\alpha,\;\; F_A =c^{A\alpha}\omega_{\alpha\beta}Z^\beta.
\label{14}
\end{equation}

Two very different situations occur.
\begin{enumerate}
\item The $X^A$ are a good set of homogenous coordinates, i.e.
the change of coordinates $z^a\rightarrow t^a\equiv X^a/X^0$ is invertible.
\item The change of coordinates $z^a\rightarrow t^a$ is not invertible.
\end{enumerate}

In the first case, one can show that the $F_A$ are the gradient of a
homogeneous function of degree two: $F_A =\partial F(X)/\partial X^A$, also
called prepotential. This is the case studied in~\cite{dW} using the $N=2$
superconformal tensor calculus. {\em
A theorem proved in~\cite{CGP} shows that in this case $N=2$ supersymmetry
cannot break to $N=1$ with zero cosmological constant}.

In the second case, the prepotential does not exist. Needless to say, this
statement depends crucially on our choice of basis eq.~(\ref{14}). This
choice is necessary to obtain a {\em local} $N=2$ Lagrangian.

This is also the case in which one {\em can} find a model with partial
supersymmetry breaking in flat space. This model was first found in
ref.~\cite{CGP2} as a singular limit of a model constructed within the
superconformal framework. In ref.~\cite{FGP2} it was realized that the model
is an example of $N=2$ supergravity without prepotential.

Specifically, the field content of the model is the minimal one necessary to
break $N=2$ to $N=1$. The hypermultiplet scalars, $b^0,b^1,b^2,b^3$,
parametrize the quaternionic manifold $SO(4,1)/SO(4)$. The two $U(1)$
symmetries, gauged by the graviphoton  and by the matter photon, act as
translations on the $b$'s:
\begin{equation}
D_\mu b^u=\partial_\mu b^u + ig\delta^{u1}A^{gph}_\mu +
i\delta^{u2}g'A^{mat}_\mu .
\label{15}
\end{equation}
The scalar of the vector multiplet parametrizes the coset manifold
$SU(1,1)/U(1)$. The K\"ahler potential is standard:
$K=-\log(z +\bar{z})$, but the symplectic basis is not:
\begin{equation}
X^0=-{1\over 2},\; X^1={i\over 2},\; F_0= iz,\; F_1=z.
\label{16}
\end{equation}
Obviously, the transformation $z \rightarrow X^0/X^1=i$ is non-invertible.

An explicit computation~\cite{FGP2} shows that the SUSY transformations of
all fermions (i.e. the gaugini $\lambda_i$, the two gravitini $\psi^i_\mu$,
and the two ``hyperini'' $\zeta^i$) are proportional to the same matrix:
\begin{equation}
\delta \, \mbox{fermions}
\propto \left(\begin{array}{ll} g-g' & 0 \\ 0 & g+g'
\end{array} \right) \left(\begin{array}{l} \eta_1 \\ \eta_2 \end{array}
\right).
\label{17}
\end{equation}
Partial breaking to $N=1$ occurs when $g=\pm g'$.

The model has a flat scalar potential $V(z,\bar{z}, b)=0$. This is not a
generic
property: it is due to our choice of the K\"ahler manifold $SU(1,1)/U(1)$.
The fields of the $N=2$ theory rearrange into $N=1$ multiplets.
Specifically, into the gravitational multiplet of $N=1$, a massive spin-3/2
multiplet, and two chiral multiplets, which are massless due to the
degeneracy of the potential.
Generalizations of this model have been given in ref.~\cite{FGI}
\section{The APT Model as a Flat Limit of $N=2$ Supergravity}
A natural question arising at this point
is whether the APT model can be
obtained from an $N=2$ supergravity in the limit of infinite Planck
mass: $M_{Pl}\rightarrow \infty$.

This is indeed the case, and the nature of the limit can be understood
quite simply, even before a detailed analysis of specific models.

First of all, the masses
of both $N=2$ gravitini should go to zero in the flat
limit $M_{Pl}\rightarrow \infty$. Indeed, a non-zero $m_{3/2}$ would
produce in the flat limit a soft {\em explicit} breaking of supersymmetry.

Moreover, the limit should produce a non-zero $C^{ij}$ in the SUSY current
algebra, or, equivalently, in eq.~(\ref{9}). The analog of eq.~(\ref{9}) in
supergravity is the so-called T-identity~\cite{CGP,FM}
\begin{eqnarray}
&& (\delta^i\lambda^I)^* Z_I^J(\phi)\delta^j\lambda_J- 3M_{Pl}^2
M^{*\, il} M^{lj}=\nonumber \\ &&  =\delta^{ij}V(\phi),
\label{18}
\end{eqnarray}
where $M^{ij}$ is the gravitino mass matrix. Eq.~(\ref{9}) is recovered in
the limit $M_{Pl}\rightarrow \infty$ when $M^{ij}\rightarrow 0$, but
$M_{Pl}M^{ij}\rightarrow $nonzero constant.

Having understood the scaling properties needed to recover the APT model, it
is relatively easy to find an explicit supergravity with the right flat
limit. The field content is as in Section 2, and the hypermultiplet scalars
still parametrize the manifold $SO(4,1)/SO(4)$. Their coupling to the
graviphoton and the matter photon is specified by the covariant
derivative~\cite{FGP}
\begin{eqnarray}
D_{\mu}b^u&=& \partial_\mu b^u + i(g_1\delta^{u3} +
g_2\delta^{u2})A^{gph}_\mu
 + \nonumber \\ && i g_3\delta^{u2}A^{mat}_\mu .
\label{19}
\end{eqnarray}

The geometry of the vector multiplet is specified by a choice of symplectic
section~\footnotemark \cite{FGP}:
\begin{eqnarray}
&& X^0={1\over \sqrt{2}},\;\; X^1={i\over \sqrt{2}} f'(z), \nonumber \\
&& F_0=-{i\over \sqrt{2}}[2f(z)-zf'(z)],\;\; F_1={z\over \sqrt{2}},
\label{20}
\end{eqnarray}
where $f(z)$ is an arbitrary analytic function, regular around $z=0$.
\footnotetext{
This section is obtained by performing the symplectic transformation
$X^1 \rightarrow -F_1$, $F_1\rightarrow X^1$ on a basis with prepotential
$F(X^0,X^1)=-i(X^0)^2f(X^1/X^0)$.}

To recover the APT model in the limit $M_{Pl}\rightarrow \infty$, we
must scale appropriately the couplings $g_1,g_2,g_3$ and the function
$f(z)$. The SUSY parameter $\eta_i$ must also be rescaled according to
its canonical dimensions, and the fermion
kinetic term must be brought in canonical form. In~\cite{FGP} it was shown
that the appropriate rescaling reads:
\begin{eqnarray}
g_1&=& {\Lambda^2\over M_{Pl}^2}\xi,\;\; g_2={\Lambda^2\over
M_{Pl}^2}e,\nonumber \\ g_3&=& 2{\Lambda\over M_{Pl}}m, \nonumber \\
f(z)&=&{1\over 2} +{\Lambda\over M_{Pl}}z +{\Lambda^2\over
M_{Pl}^2}\phi(z) + \nonumber \\ &&
+{\cal O}(\Lambda^3/ M_{Pl}^3),
\nonumber \\ \lambda_i&\rightarrow & (M_{Pl}\Lambda^2)^{-1/2}\lambda_i,
\;\; \zeta^i\rightarrow M_{Pl}^{-3/2}\zeta^i,\nonumber \\
\psi_{i\, \mu}&\rightarrow & M_{Pl}^{-3/2}\psi_{i\, \mu} ,\;\;
\eta_i\rightarrow M_{Pl}^{1/2}\eta_i.
\label{21}
\end{eqnarray}
The limit is taken keeping the scale $\Lambda$,the dimensionless
parameters $\xi,e,m$, and the function $\phi(z)$ constant.
In this limit the entire gravitational
multiplet, as well as the hypermultiplet, decouple from the vector multiplet.
The gravitational multiplet and the hypermultiplet become a hidden sector,
whose interactions with the vector multiplet (the ``observable sector'') are
suppressed by inverse powers of $M_{Pl}$. For instance, the kinetic term
of the hypermultiplet scalars reads $M_{Pl}^2 2^{-1} b_0^{-2}\partial_\mu
b^u \partial^\mu b^u$. In the limit $M_{Pl}\rightarrow \infty$, the
$b^u$'s do not fluctuate and appear as coupling constants in the low-energy
action~\footnotemark.
\footnotetext{This is what
happens in string theory with the $S$ modulus.}
By defining ${\cal F}(z)=z^2 -2i\phi(z)$, one finds that, in the flat limit,
the observable sector of our model becomes identical with the APT model, with
parameters given in eq.~(\ref{5a}).
In particular, using the same definition of $\tau(z)$ as in Section 1,
the SUSY transformation laws of the gaugini on a Lorentz-invariant background
read:
\begin{eqnarray}
\delta\lambda_i&=&X_{ij}\eta^j +
{\cal O}(\Lambda^3/ M_{Pl}), \nonumber \\
X_{ij}&=& \sqrt{2}{\Lambda^2\over b^0 \tau_2(z)}\{
\xi \sigma^1_{ij} + i[e + m\tau(z)]\delta_{ij}\}\eta^j .\nonumber \\ &&
\label{22}
\end{eqnarray}
At $\tau_1(z)=-e/m$, $\tau_2(z)=\xi/m$, the matrix $X_{ij}$ has a
nondegenerate zero eigenvalue. As expected from the general considerations
outlined before, there the scalar potential has a stable minimum in $z$.
At the minimum, the potential reads
\begin{equation}
V|_{min}=-{\Lambda^4\over 2(b^0)^2}[(2m-\xi)^2 +e^2]
\label{23}
\end{equation}
\section{Conclusions}
We have seen that global extended
supersymmetry can indeed undergo partial spontaneous
breaking, and we showed that what makes it possible, is a field-independent
modification of the SUSY current algebra, which does not modifies the SUSY
algebra {\em on local, gauge-invariant operators}.
We have also shown that
partial breaking of supersymmetry at zero cosmological constant can occur
in extended supergravity, by studying the $N=2$ case in some details.

Finally, we have shown that the APT model of global
extended supersymmetry with partial SUSY breaking can be obtained as a flat
limit of a supergravity model. In our example, the supergravity model
contains extra fields that decouple only in the flat limit. Supersymmetry is
broken to $N=1$ in the observable (vector-multiplet) sector. As shown
in~\cite{FGP}, the hidden sector generically breaks $N=2$ to $N=0$, in which
case the scalar potential has no stationary point in $b^0$ for large
but finite $M_{Pl}$.
For the special choice of parameters $e=0$, $\xi=2m$, the hidden sector
too
preserves $N=1$ supersymmetry, and the scalar potential has a flat direction
in $b^0$, as it can be verified by inspection of eq.~(\ref{23}).

\end{document}